# Agglomerative Hierarchical Clustering Analysis of co/multi-morbidities

Shatrunjai P. Singh[1], Swagata Karkare[2], Sudhir M. Baswan[3] and Vijendra P. Singh[4]

[1]Lindner College of Business, University of Cincinnati, Ohio, USA; [2]School of Public Health, Boston University, Boston, MA, USA; [3]James L. Winkle College of Pharmacy, University of Cincinnati, Ohio, USA; [4]Department of Internal Medicine, Baptist Easley Hospital, Easley, SC, USA.

1. **ABSTRACT:**

Although co/multi-morbidities are associated with significant increase in mortality, the lack of appropriate quantitative exploratory techniques often impede their analysis. In the current study, we study the clustering of multimorbid patients in the Texas patient population. To this end we employ agglomerative hierarchical clustering to find clusters within the patient population. The analysis revealed the presence of nine distinct, clinically relevant clusters of co/multi-morbidities within the study population of interest. This technique provides a quantitative exploratory analysis of the co/multi-morbidities present in a specific population.

2. **INTRODUCTION:**

One in four American have two (comorbid) or more (multimorbid) chronic conditions[1]. Projections estimate that more than 81 million Americans will suffer from these multi-morbidities by 2020.[2] Further, new data suggest that routine clinical procedures when applied to patients with multimorbidity can lead to unintended adverse events[3]. Thus there is a critical need to identify patients with these patients with complex co and multi-morbidities to enable proper support and intervention. Previous research to classify subgroups of composite patients with co/multi-morbidities have conventionally depend upon complex multivariate regression techniques[4,5]. Although, newer supervised and unsupervised machine learning algorithms have been successfully adopted in many spheres of biomedical data analysis, they have not been tried on classifying co/multi-morbidities in patient data[6,7].

Clustering, an unsupervised machine learning technique that aims to grouping analogous entities into one cluster and partitioning the dissimilar objects into another cluster[8–10]. A cluster is defined as a subset of similar objects, defined by certain parameters, within a larger set. The threshold definition of similarity cutoff is often subjective and is usually determined by the study design. With respect to medical conditions, a multi-morbidity cluster can be defined as an unsupervised technique to find patients with similar medical conditions. In the current study, we use correlational clustering analysis to find the key group of diseases present in the population. We further employ hierarchical clustering analysis on patients from the Texas health care patient data to describe inherent patterns in clusters of multimorbid patients. The described clustering approaches identifies cohorts of multi-morbidities and presents opportunities for better management of these patients.

## 3. METHODS:

### 3.1 STUDY POPULATION

We used open access de-identified aggregate data provided by the Texas Department of State Health Services (http://healthdata.dshs.texas.gov/Home) to conduct this analysis. Inpatient and Outpatient datasets were combined to generate a composite dataset consisting of more than 15,000 data points and the inpatient procedure code was used to identify different clinical conditions. The training cohort consisted of patients who 21 years or more as of January 1, 2015, with two (comorbidities) or more (multimorbidities) identified by inpatient procedure code on first examination. Members with admits to hospice, a long term care facility, or with a pregnancy reported in the last 3 months were excluded from the study. After exclusions our final study population was 13,920 patients. We isolated the list of 75 most common conditions reported by the Center for Disease Control, USA and used it to further filter out input dataset[11]. A literature search was also performed to further identify conditions which could be included in the study based on the general Texas population, our specific study cohort and disease with relatively high prevalence. Identification of conditions within cohort members were based on an outpatient data cross-referenced to *International Classification of Diseases, Tenth Revision (ICD-10)* diagnosis and procedure codes in 2015[12]

### 3.2 EXPLORATORY DATA ANALYSIS AND FEATURE GENERATION

Microsoft SQL Server (version 2012) was used to extract, transform, load and query the dataset. Binary outcome variables were created for the selected conditions. Age, gender, income and other demographic variables were also included in the input dataset. Input variables were scanned for outliers. Imputations to median/mode were performed for the non-binary continuous/categorical variables. We observed less than 2% imputations overall in the dataset. A non-zero variance analysis was performed on the binary disease variables and variables with less than 2% variance were further excluded from the analysis.

### 3.3 STATISTICAL ANALYSIS

All statistical analysis was performed using R-statistical software (Version 0.98.109). Significance testing for normal, non-normal and binary data was performed as described previously by the authors in other studies [13–24]. De-identified data and the statistical analysis R code used was uploaded to an online repository.

### 3.4 CORRELATION ALGORITHM

Correlation clustering involves the creation of a weighted matrix X=(P,E), such that the edge weight specifies the similarity ($+^{ve}$ edge weight) or dissimilarity ($-^{ve}$ edge weight). The goal is to find an optimal cluster which maximizes similarity or minimizes dissimilarity[8]. The method of minimizing disagreement was chosen for the current study based on the characteristics of the input data. A spearman's correlation matrix was generated for binary variables created from different input conditions. A k-means clustering algorithm based on Jaccard's distance was created and analyzed. The optimal cluster number was chosen to minimize the goodness of fit criterion. The cluster of conditions was analyzed for similarity of condition based on origin, organ system and patient demographic.

### 3.5 CLUSTERING ALGORITHM

An agglomerative hierarchical clustering (AHC) algorithm with a bottom up approach was used to separate clinically appropriate clusters within the study population. The bottom up reproach to

AHC initiates with each member starting at an isolated cluster, followed by serial merging of similar members to form similarity clusters until only once cluster remains. After the clustering procedure terminates, subject matter expertise, clinical relevance and study design criterion are used to select a cutoff/threshold which produces the final clusters. The process can be visualized using dendrograms. We used Ward's method along with Gower's distance matrix for similarity calculations as it has shown to be more reliable for mixed data with a preponderance of weighted binary data (like condition related binary variables)[25].

### 3.6 FIGURE PREPARATION

The results from R-software were exported into csv files which were imported into Tableau (version 8.0) or Microsoft Excel (version 2013) which were then used to create graphs and visualizations. Tables were created in Microsoft Word (version 2013).

## 4. RESULTS:

### 4.1 CORRELATION ANALYSIS REVEALS EXPECTED CLUSTER OF MAJOR CONDITIONS IN THE PATIENT POPULATION

A spearman's correlation analysis was performed on the dataset of more than 70 different conditions to check for the correlation between different conditions in the population. The resulting correlation matrix was further clustered to produce grouping of similar conditions with a correlation coefficient cutoff greater than 0.60 (Figure 1).

We identified 5 broad cluster of multi-morbid conditions: (1) diseases overrepresented in the female population including menopause, Chronic Thyroid disorder and Osteoporosis; (2) Neurological and Psychiatric conditions including substance related psychiatric conditions, epilepsy, dementia and depression; (3) Disease related metabolic syndrome including hypertension, lower back pain, hyperlipidemia, obesity and diabetes mellitus; (4) conditions related to the cardiovascular system including heart failure, ischemic heart disease, peripheral artery disease and cerebrovascular disease; (5) conditions of the eye including cataract and glaucoma. The clusters identified were homogenous and overall had low demographic variance. A radial dendrogram was created to further visualize the similarity of conditions within the population (Figure 2).

## 4.2 CLUSTERING ANALYSIS REVEALS 9 BROAD CLUSTER OF MULTI-MORBIDITY PATIENTS IN THE POPULATION

Agglomerative Hierarchical Clustering revealed 9 broad clusters in the population data of 13,920 patients (Figure 3). The average age of patients was 54.9 years and contained 50.1 % males and 49.9 % females. Descriptive statistics revealed clinical homogeneity within the clusters (Table 1). The clusters (numbered randomly) were divergent based on the mix of co/multi-morbidities observed and age/gender demographics. Clinically relevant summarization showed the presence of distinct clusters with a high proportion of patients with (Table 2) : cancer (cluster 1), musculoskeletal diseases (cluster 2), substance abuse (cluster 3), female population with arthritis and post-menopausal conditions (cluster 4), metabolic syndrome related conditions (cluster 5), thyroid related conditions (cluster 6), females with migraines and depression (cluster 7), elderly population with multiple conditions (cluster 8) and a diabetes cohort (cluster 9). Income, medical utilization, inpatient visits were also calculated but are not shown in the current analysis.

## 5. DISCUSSION:

In the current paper we use a clustering approach to identify groups of patient with similar multimorbidity conditions in the Texas population. The clusters identified were homogenous and clinically relevant with actionable insights. We present a fast and easy approach to explore patient data for faster insight generation.

Co/multi-morbidities are illnesses that coexist with a condition of interest and often lead to delayed or misdiagnosis and have been shown to increase mortality in multiple populations[26]. They are a major source of economic burden on the healthcare systems with multimorbid patients experiencing worse health, economic and social outcomes compared to mono-condition patients. Indeed, It has been shown that having multiple health conditions significantly increases the probability of reporting a diminished quality of life[26,26,27]. Comorbidity indices are frequently used to summarize the overall health of a population but offer suffer from errors of manual data curation[28,29]. Our analysis provides a quantitative, data driven approach to exploring multimorbid patient data with the possibility of real time analysis.

Clustering on the Texas patient health data to isolate multimorbidity patients revealed nine well defined clusters. An analysis of the most prevalent conditions in every cluster revealed broad groupings within each cluster (Table 2). Our first cluster was also the largest with 4,532 patients. It contained middle aged patients (median age 53.5 years) and had a slightly higher ratio of males (56%) compared to females (44%). The cluster was characterized by the highest incidence of cancers of different organ systems including colorectal cancer (2.8%), prostate cancer (3.9%), ovarian cancer (0.6%), Multiple myeloma (1.5%), malignant melanoma (1.8%), pancreatic cancer (0.5%), esophageal cancer (0.4%), stomach cancer (0.4%), skin cancer (7.2%), oral cancer (0.8%) and other cancers (8.9%). Interestingly, this cluster also had the highest incidences of kidney stones (7.7%), inflammatory bowel disease (8.6%) and sickle cell anemia (0.2%) indicating a potential causal role of these conditions in certain cancers. Indeed, a 2015 meta-analysis revealed an increased risk for kidney stone formation and renal cell carcinoma[30]. Further corroborating cluster derived corollary relationship, inflammatory bowel disease patients are known to be at an increased risk for colorectal cancer and have been recently identified to be a risk factor for oral cancer[31,32].

Our second cluster had a higher ratio of older females (53% females with median age of 59.6 years) and had the highest incidences of conditions like osteoarthritis (69%) and lower back pain (71%). Interestingly, this cluster in general tends to be more expensive compared to other clusters (results not shown). Cluster three was our youngest cluster (median age 42.9 years), with more males compared to females (64% males). This cluster contained the highest percentage of members with substance abuse (80%) and its related disorders including hepatitis (15%), pancreatitis (16%), neurosis (1.1%), and psychoses (9.3%). Remarkably, in this cohort we also saw the highest rate of post-partum neurosis disorders raising the possibility of substance abuse or addiction post-partum. Indeed it has been suggested that pharmacological agents used to treat post-partum depression often lead to long term addiction and need more regulation[33,34].

Cluster 4 contained a high proportion of middle aged females (median age 56.1 years; 62.5 % females) with a relatively high proportion reporting menopause (17.2%). Further, this cluster also reported the highest incidences for conditions like rheumatoid arthritis (9.4%) and fibromyalgia (8.8%). This correlative evidence further backs the previous similar observations of links between fibromyalgia, rheumatoid arthritis and menopause[35,36]. Martinez-Jauand and colleagues, have previously shown that an early menopause can reduce estrogen exposure and this causes an increased sensitivity to pain magnifying fibromyalgia symptoms[35]. Cluster 5, although relatively smaller in size (895 patients) contained a very high proportion of patients with metabolic syndrome (95%) patients. As expected this cluster had the highest rates of hypertension (98%) and obesity (61%). This cluster also contained a high proportion of cervical cancer patients (0.4%) and this link has previously been demonstrated in other populations[37,38]. Cluster 6, was our smallest cluster cohort (765 patients) with the highest proportion of females (77.3% females) had the highest incidences of osteoporosis (22.4%), chronic thyroid disorders (86.4%), chronic fatigue syndrome (1.2%). A common theme related to these conditions is the interleukin-6 pathway, dysregulations of which are known to play a central role in osteoporosis, thyroid disorders and neck cancer[39–42]. Our results suggest that this cytokine molecular pathway may be responsible for more disorders than previously identified.

Cluster 7 also had a high proportion of young females (59% females; median age 49.1 years) who reported a high proportion of neurological and psychological related disorders including psychosis (2.8%), depression (48%), migraines/headaches (35.4%), epilepsy (12.1%), and eating disorders (0.6%). This group also had the highest proportion of fertility issues (0.7%) and gave birth to babies with low birth weight (0.1%). We also observed this group reporting an increased incidence of having a disrupted childhood (0.6%), posing a possible origin of these psychological issues. Similar multi-morbidity associations have extensively been studied in childhood post-traumatic stress disorders[43–45]. Cluster 8, our oldest cluster (median age 60.5 years) with the highest proportion of males (66.8% males) suffered from a combination of cardiovascular and respiratory diseases commonly seen in the elderly population. This cluster had a high incidence rates for heart failure (49%), cerebrovascular disease (80.9%), COPD (16.7%), congenital heart disease (1.7%) and ventricular arrhythmia (14.2%). Further, we also highest incidences of Parkinson's (0.7%) and dementia (2.7%) as expected from this cluster. This cohort of patients were seen to have the highest frequency of in-patient visits and the highest total cost associated with them (data not shown). Our final cluster 9, was dominated by middle aged males (61.8% males; median age 59.2 years) with the highest incidence of diabetes mellitus (68.8%). We also saw highest incidences of diabetes related chronic disorders like chronic renal failure (29.5%), cataract (20.8%) and glaucoma (20.6%). Relationships between these conditions have been extensively reported[46–48].

Overall our clustering approach has identified cohorts of patients with similar multimorbid diseases with actionable insights which can be used to reduce disease incidence and overall cost.

## 7. FIGURES AND TABLES LEGENDS

**Figure 1:** A matrix showing the results of spearman's correlation analysis followed by k-means clustering on the results. Conditions are represented on the top and left panel. The size of the circles depict the strength of correlation between diseases. An additional color coding of the spearman's correlation coefficient (red to blue signifies correlation from -1 to +1) was added to increase interpretability. Clustering analysis using the wards method reveals five clusters of clinically related conditions shown in the dotted red box.

**Figure 2:** A closed radial dendrogram shows the structure of different conditions as inferred from the data. Conditions which frequently co-occur in patients share a common node (branch) in the dendrogram.

**Figure 3:** Results from hierarchical agglomerative clustering reveal nine distinct clusters. Clustering was performed using Ward's method with Gower's distance and a threshold (h=27; shown as a dotted line) was used to isolate 9 clusters.

**Table 1:** Summaries of disease distributions in different clusters. This table shows cluster summaries for age (median; years), male and female composition (%), and the proportion of people identified with different medical conditions (%; number members in cluster with the disease/ total number of members with the disease), for the nine clusters.

**Table 2:** A summary of clinical findings from each cluster.

# 8. FIGURES AND TABLES:

Figure 1:

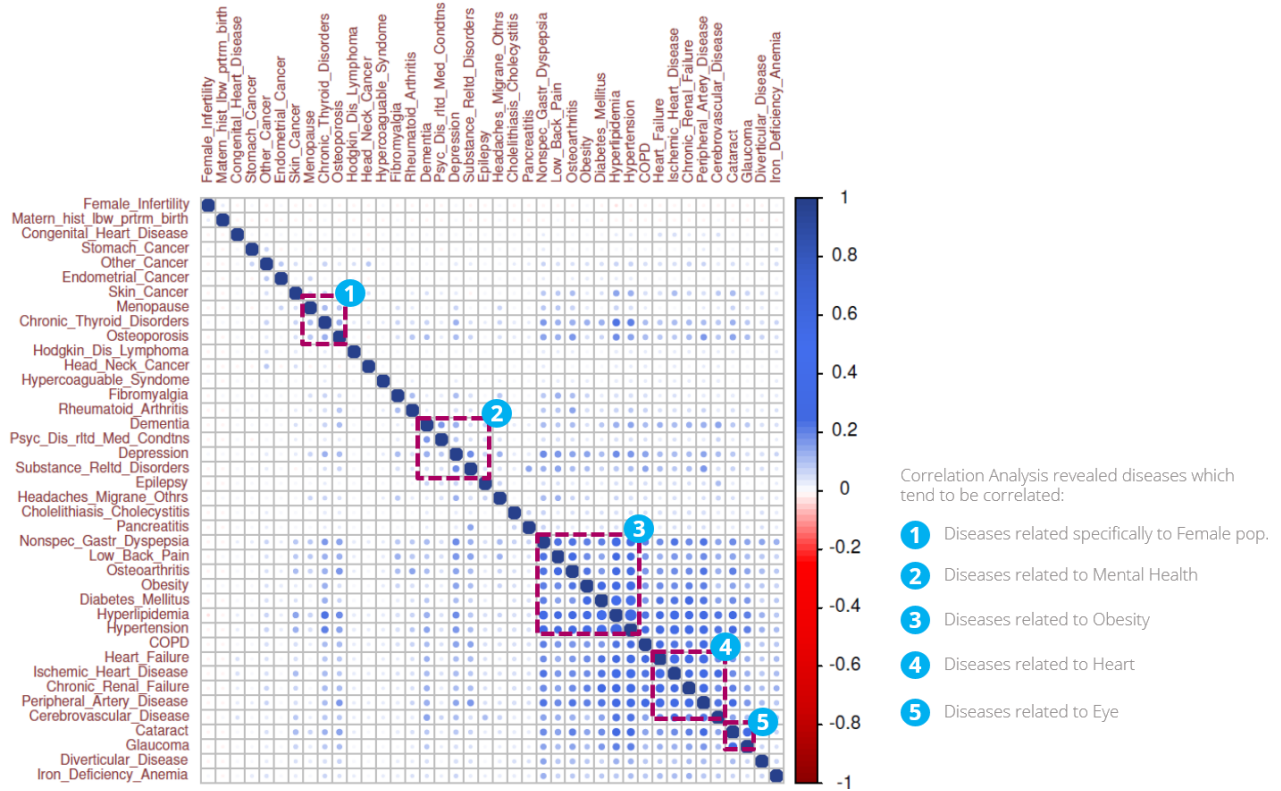

Figure 2:

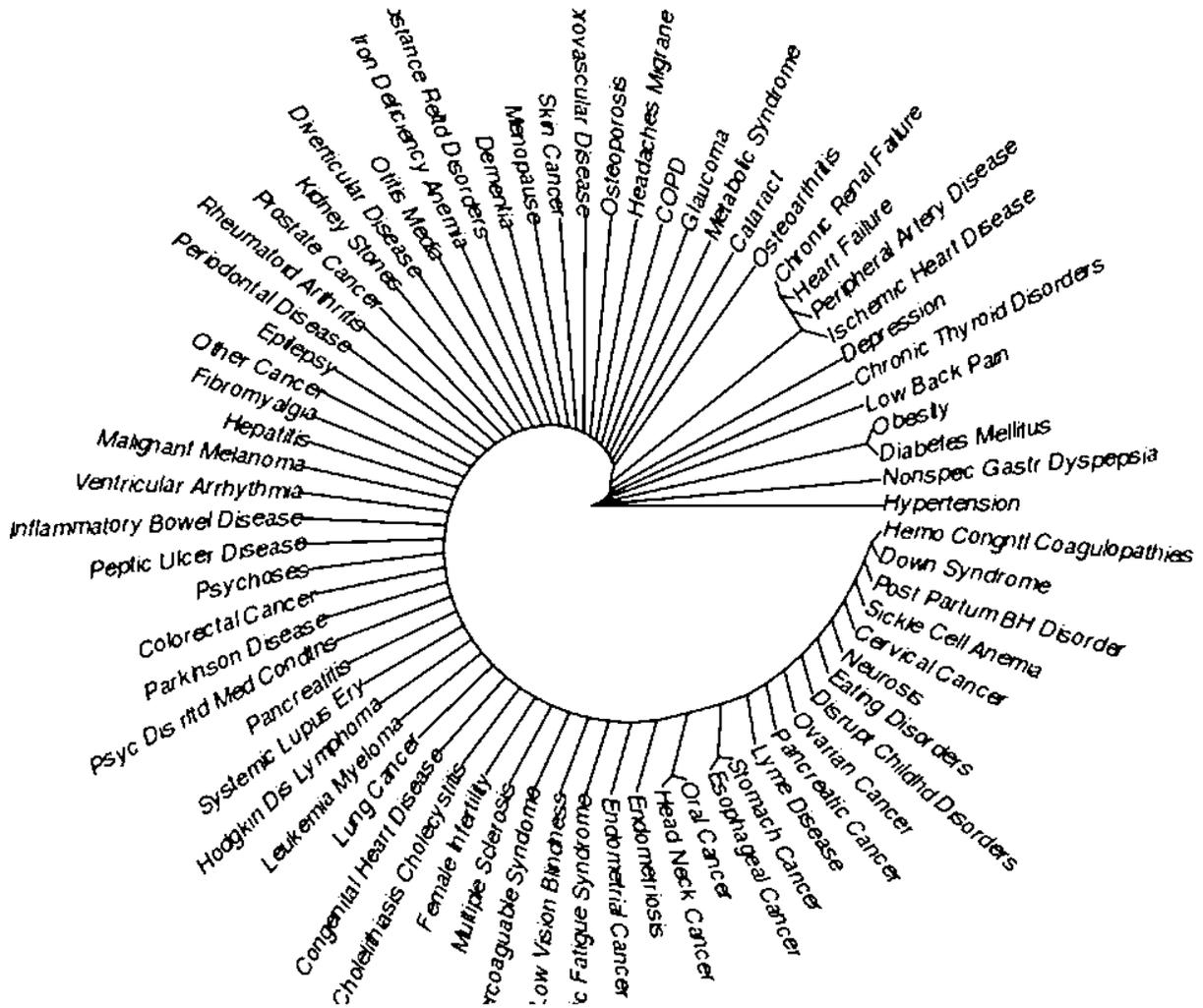

Figure 3:

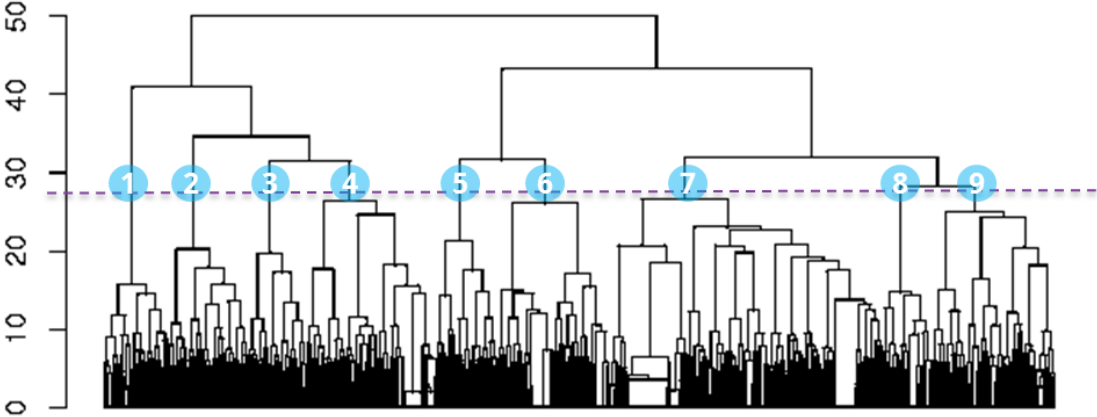

Table 1:

| Cluster Number | 1 | 2 | 3 | 4 | 5 | 6 | 7 | 8 | 9 |
|---|---|---|---|---|---|---|---|---|---|
| **Size** | 4532 | 976 | 733 | 1564 | 895 | 765 | 1655 | 1101 | 1699 |
| **Age** | 33.9 | 66.5 | 58.3 | 51 | 46.8 | 58 | 74.6 | 69 | 54.5 |
| **Males** | 52.6 | 59.8 | 48.1 | 49.2 | 34.9 | 23.9 | 42.2 | 44.2 | 44.3 |
| **Females** | 47.4 | 40.2 | 51.9 | 50.8 | 65.1 | 76.1 | 57.8 | 55.8 | 55.7 |
| **Diabetes Mellitus** | 0.1 | 55.8 | 8.6 | 11.5 | 9.6 | 12.7 | 60.9 | 17.7 | 16.1 |
| **Heart Failure** | 0 | 17.5 | 0.7 | 0.3 | 1.9 | 1.8 | 41.8 | 5.5 | 3.7 |
| **COPD** | 0 | 6.8 | 1.8 | 1.2 | 3.2 | 1.1 | 35.9 | 23.4 | 2.8 |
| **Ischemic Heart Disease** | 0 | 35.7 | 1.2 | 3.1 | 1.9 | 3.6 | 51.6 | 6.7 | 2.6 |
| **Otitis Media** | 2 | 0.6 | 0.7 | 0.9 | 0.3 | 1.6 | 1.8 | 0.7 | 0 |
| **Peptic Ulcer Disease** | 0 | 0.5 | 0.3 | 0 | 0.5 | 0.2 | 3.1 | 0.6 | 2 |
| **Hypertension** | 0.2 | 80.1 | 58.4 | 35 | 34.9 | 33.3 | 96.9 | 64 | 38.4 |
| **Epilepsy** | 0 | 0.5 | 2 | 0 | 2.3 | 1.1 | 2.6 | 1 | 0.2 |
| **Chronic Thyroid Disorders** | 0.2 | 13.9 | 7.6 | 7.4 | 8.4 | 99.1 | 29 | 15.6 | 23.8 |
| **Depression** | 0.1 | 6.8 | 4.5 | 8.4 | 64.1 | 15.2 | 30 | 11.3 | 10.5 |
| **Cholelithiasis Cholecystitis** | 0.2 | 0.6 | 0.3 | 0.6 | 0.3 | 0.4 | 1.6 | 0.6 | 0.2 |
| **Iron Deficiency Anemia** | 0.1 | 1.3 | 3.3 | 1.2 | 0.8 | 0.4 | 12.2 | 0.4 | 2.4 |
| **Osteoarthritis** | 0.1 | 5.4 | 3.2 | 0.6 | 3.1 | 9.8 | 37.6 | 49.6 | 5.5 |
| **Rheumatoid Arthritis** | 0.2 | 1 | 0.1 | 0.3 | 0.6 | 1.8 | 4 | 2.9 | 3.1 |
| **Colorectal Cancer** | 0.1 | 0.6 | 0.6 | 0.6 | 0.2 | 0.4 | 2.8 | 0.7 | 0.7 |
| **Lung Cancer** | 0 | 0.1 | 0.3 | 0 | 0.2 | 0.4 | 1.2 | 1.1 | 0 |
| **Hyperlipidemia** | 3.7 | 74 | 43.1 | 29.1 | 34.3 | 44 | 89.1 | 61.6 | 42.4 |
| **Cerebrovascular Disease** | 0 | 9.4 | 5.1 | 0.3 | 1.9 | 1.8 | 26.6 | 5.5 | 3.7 |
| **Headaches & Migraines** | 0.3 | 1.2 | 3.6 | 4 | 39 | 1.3 | 8.2 | 1.9 | 5 |
| **Chronic Renal Failure** | 0 | 19.7 | 2.5 | 0.3 | 1.3 | 2 | 47.5 | 6.6 | 1.7 |
| **Kidney Stones** | 0 | 2 | 3.9 | 1.5 | 0.5 | 0.4 | 3.1 | 2.2 | 1.3 |
| **Diverticular Disease** | 0 | 1.2 | 3.9 | 0 | 1.3 | 0.7 | 9.7 | 2.7 | 1.7 |
| **Low Back Pain** | 0.1 | 9.8 | 6.5 | 99.7 | 14 | 6.7 | 49.4 | 25.4 | 20.5 |
| **Nonspecific Gastric Dyspepsia** | 0.4 | 15.9 | 13.2 | 12.1 | 22.2 | 5.6 | 69.1 | 28.8 | 97.4 |
| **Sickle Cell Anemia** | 0 | 0 | 0 | 0.9 | 0 | 0 | 0.1 | 0.1 | 0.4 |
| **Multiple Sclerosis** | 0.2 | 0 | 0.3 | 0 | 0.3 | 0 | 0.1 | 0.1 | 0.2 |
| **Inflammatory Bowel Disease** | 0.3 | 0.7 | 0.6 | 0.3 | 0.2 | 0 | 1 | 0.4 | 0.7 |
| **Hemo Congenital Coagulopathies** | 0 | 0 | 0.2 | 0 | 0 | 0 | 0.1 | 0.1 | 0.2 |
| **Systemic Lupus** | 0 | 0.1 | 0.6 | 0.3 | 0.3 | 0 | 0.9 | 0.4 | 0.4 |
| **Prostate Cancer** | 0 | 3 | 3.7 | 0.9 | 0.5 | 0.9 | 7.2 | 2.6 | 0.4 |
| **Ovarian Cancer** | 0 | 0.1 | 0 | 0 | 0.2 | 0.2 | 0.4 | 0.4 | 0.2 |
| **Endometrial Cancer** | 0 | 0.1 | 0 | 0.3 | 0.2 | 0.2 | 0.7 | 0.6 | 0 |
| **Cervical Cancer** | 0.1 | 0 | 0.1 | 0 | 0 | 0.2 | 0 | 0.1 | 0 |
| **Hodgkin Dis Lymphoma** | 0.1 | 0.5 | 0.4 | 0.3 | 0 | 0 | 0.7 | 0.5 | 0.4 |
| **Leukemia Myeloma** | 0.1 | 0.4 | 0.3 | 0.3 | 0.2 | 0 | 1 | 0.2 | 0.4 |

| Condition | | | | | | | | | |
|---|---|---|---|---|---|---|---|---|---|
| Malignant Melanoma | 0 | 0.5 | 1.9 | 0 | 0.2 | 0.2 | 1.5 | 0.4 | 1.3 |
| Head Neck Cancer | 0 | 0.1 | 0.3 | 0 | 0.2 | 0.2 | 0.3 | 0 | 0.4 |
| Esophageal Cancer | 0 | 0.1 | 0.1 | 0 | 0 | 0 | 0 | 0.2 | 0.2 |
| Stomach Cancer | 0 | 0 | 0 | 0 | 0 | 0 | 0.3 | 0.1 | 0.2 |
| Pancreatic Cancer | 0 | 0.4 | 0 | 0 | 0.2 | 0 | 0.4 | 0 | 0 |
| Pancreatitis | 0 | 0.6 | 0.1 | 0 | 1.3 | 0 | 2.4 | 0.5 | 0.4 |
| Hepatitis | 0 | 0.1 | 1.8 | 0 | 1 | 0 | 1.5 | 0.9 | 1.1 |
| Peripheral Artery Disease | 0 | 22.2 | 2.1 | 0 | 1.8 | 0.4 | 50 | 11.5 | 2.4 |
| Endometriosis | 0.1 | 0 | 0.1 | 0 | 0.2 | 0.2 | 0 | 0 | 0 |
| Ventricular Arrhythmia | 0 | 2.8 | 0.3 | 0.9 | 0.3 | 0 | 6.5 | 0.4 | 0.2 |
| Lyme Disease | 0 | 0 | 0.1 | 0 | 0 | 0 | 0 | 0.1 | 0 |
| Female Infertility | 0.3 | 0 | 0 | 0.6 | 0.5 | 0.2 | 0 | 0 | 0.2 |
| Menopause | 0 | 0.8 | 10.4 | 2.8 | 1.9 | 3.3 | 4.9 | 4.5 | 2.2 |
| Glaucoma | 0.1 | 7.3 | 2.5 | 0.9 | 1.1 | 2 | 28.2 | 29.8 | 2.8 |
| Low Vision Blindness | 0 | 0.5 | 0.2 | 0.3 | 0.2 | 0 | 1.2 | 1.2 | 0.6 |
| Cataract | 0.1 | 10.2 | 4 | 0.6 | 2.1 | 3.8 | 40.3 | 36.2 | 9.4 |
| Other Cancer | 0 | 1.3 | 1.1 | 0.3 | 0.6 | 3.1 | 4.6 | 1 | 2 |
| Dementia | 0 | 2.4 | 3 | 0.3 | 0.8 | 0 | 12.8 | 2.1 | 0.4 |
| Osteoporosis | 0.1 | 4.3 | 7.7 | 4.6 | 0.6 | 0.9 | 17.2 | 8.2 | 6.1 |
| Obesity | 0.2 | 23.4 | 20.3 | 8.4 | 15.7 | 16.5 | 52.1 | 13.6 | 5 |
| Oral Cancer | 0 | 0 | 0 | 0 | 0 | 0 | 0.1 | 0 | 0 |
| Cystic Fibrosis | 0 | 0 | 0.1 | 0 | 0 | 0 | 0 | 0 | 0 |
| Neurosis | 0 | 0 | 0.1 | 0.3 | 0.5 | 0 | 0.3 | 0 | 0 |
| Psychoses | 0 | 0.2 | 0.1 | 0.3 | 3.4 | 0.7 | 2.1 | 0.1 | 0 |
| Eating Disorders | 0 | 0 | 0.1 | 0 | 0.3 | 0 | 0 | 0 | 0 |
| Disrupt Childhd Disorders | 0.2 | 0 | 0.1 | 0 | 0 | 0 | 0 | 0 | 0 |
| Substance Reltd Disorders | 0 | 1.4 | 0.5 | 0.3 | 9.1 | 0.7 | 12.5 | 3 | 0.6 |
| Skin Cancer | 0.1 | 5.3 | 10 | 0.6 | 0.5 | 0.9 | 11.2 | 5 | 1.3 |
| Congenital Heart Disease | 0.2 | 0.5 | 0 | 0 | 0 | 0 | 0.7 | 0.5 | 0 |
| Periodontal Disease | 0 | 1 | 5 | 0.3 | 0.2 | 0 | 0.3 | 0.9 | 0.2 |
| Chronic Fatigue Syndrome | 0.1 | 0.2 | 0.2 | 0 | 0.3 | 0 | 0.1 | 0.4 | 0.6 |
| Fibromyalgia | 0 | 0.6 | 0.2 | 1.2 | 4.7 | 0.9 | 3.4 | 2.1 | 2.2 |
| Parkinson Disease | 0.2 | 1 | 0.6 | 0.9 | 0.2 | 0.2 | 1.3 | 0.4 | 0.2 |
| Hypercoaguable Syndome | 0.1 | 0.2 | 0.2 | 0 | 0.3 | 0 | 0.6 | 0.2 | 0.2 |
| Post Partum BH Disorder | 0 | 0 | 0.1 | 0 | 0 | 0 | 0 | 0 | 0 |
| Matern low birth weight | 0 | 0 | 0 | 0 | 0 | 0 | 0 | 0 | 0 |
| Metabolic Syndrome | 0 | 3.5 | 12.8 | 8.4 | 8.3 | 7.4 | 6.6 | 5.9 | 3.1 |
| Psyc. Disorder | 0 | 0.4 | 1.6 | 0.3 | 0.5 | 0.2 | 2.1 | 0.6 | 0.2 |

Table 2:

| Cluster 1 | Cluster 2 | Cluster 3 | Cluster 4 | Cluster 5 | Cluster 6 | Cluster 7 | Cluster 8 | Cluster 9 |
|---|---|---|---|---|---|---|---|---|
| Cancer | Musculoskeletal disorders | Substance Abuse | Menopause & Rheumatoid Arthritis | Obesity and Hypertension | Thyroid and Osteoporosis | Migraine and Depression | Elderly population with multiple conditions | Diabetes |